# Exploring the potential of GeTe for the application in Thermophotovoltaic (TPV) cell

Ahnaf Tahmid Abir[1], Bipanko Kumar Mondal[1,2] and Jaker Hossain[1*]

[1]Solar Energy Laboratory, Department of Electrical and Electronic Engineering, University of Rajshahi, Rajshahi 6205, Bangladesh.

[2]Department of Electrical & Electronic Engineering, Pundra University of Science & Technology, Bogura, Bogura 5800, Bangladesh.

**Abstract**

Germanium telluride (GeTe) having a direct bandgap of 0.6 eV has mainly been in phase change memory and thermoelectric power generation. In this article, we study the electronic structure of the GeTe by first-principles calculations. The theoretical direct bandgap of GeTe was found to be 0.69 eV which is very close to the experimental value. Then, we demonstrate a single-junction GeTe thermophotovoltaic (TPV) cell based on device transport model with np structure. The device was optimized for the higher performance of the TPV cell. The GeTe TPV cell exhibited an efficiency of 7.9% with $J_{SC}$=16.16 A/cm$^2$, $V_{OC}$=0.360 V and FF=75.51%, respectively. These results indicate that GeTe could be a promising material for the fabrication of efficient TPV cell.

**Keywords:** GeTe, DFT, band structure, pn junction, TPV cell, device transport model.

## 1. Introduction

A thermophotovoltaic (TPV) cell is a semiconductor pn junction that converts thermal radiation of a high temperature source into electrical energy. The typical temperature range of thermal sources i.e. black bodies (BB) is around 1000-2000K, therefore the band gap of TPV materials should be in the range of near infrared [1]. Huge amount of heat energy is injected into the environment as waste heat in high energy consumption industries in the form of hot exhaust gases, cooling water, and heat lost from the surfaces of hot equipment around the earth. A TPV cell can be effectively used for harvesting this heat for improved energy usage and reduction of carbon emissions [2]. However, the TPV literature indicates that TPV cells are mainly comprised of III-V or II-VI narrow bandgap semiconductors, particularly cells based on indium arsenide (InAs), indium antimonide (InSb), lead selenide (PbSe), gallium antimonide (GaSb), indium gallium arsenide (InGaAs), indium gallium arsenide antimonide (InGaAsSb), indium arsenide antimonide phosphide (InAsSbP), and indium gallium arsenide antimonide phosphide (InGaAsSbP). The experimental works show that the efficiencies of these TPV cells are in the range of 1-11% although the theoretical efficiency limit is ~24% [3-5]. Moreover, the widely used TPV cells based on GaSb, Ge, InGaAs, GaSb and GaInAsSb semiconductors are relatively wider bandgap materials for waste heat recovery.

The main focus of this work is to explore the potential of a narrow band gap Germanium telluride (GeTe) semiconductor in TPV application. GeTe is a IV-VI semiconductor having a direct bandgap of 0.6 eV and it is useful at the intermediate temperature due to its high thermoelectric (TE) performance [6-7]. GeTe has fascinated for having various properties such as thermoelectric, ferroelectric and phase-change material during the period of late 1960. Moreover, it also bears some other qualities like limited power consumption, retention of data, storing capacity and enormous high speed that makes GeTe a trusted candidate for volatile memory techniques. This qualities also lead GeTe as a prime element for both non-volatile memory (NVM) and Phase change random access memory (PCRAM) devices [8-10]. Faster crystallization rate and larger resistance prove this binary chalcogenide material's suitability for a Phase change memory (PCM) [11]. However, to the best of our knowledge, there is few reports of using GeTe in TPV cells.

Generally, the GeTe thin film is prepared from either thermal evaporation or by sputtering process in the form of amorphous or crystal [12]. Several researchers worked on the structure of GeTe and found that it is rhombohedral with an angle around 90° below 127K [13]. Recent reports indicate that single crystalline GeTe film can be grown on KCl substrate and in phase transformation process they form extended twins of {l00} and {110} type [14]. The GeTe thin film can be deposited on NaCl substrate by evaporation process below the temperature of 573K but above this rated temperature re-evaporation takes place which prevents the growth of the PCM material on salt. However, this phase changing material can also be grown on a silicon substrate at room temperature by sputtering process which is common nowadays [15]. There is also a report on deposition of GeTe on silicon and glass substrate by ALD method [16].

Pristine GeTe shows a high p-type conductivity with a hole concentration of $\sim 10^{21}$ cm$^{-3}$ owing to its intrinsic Ge vacancies leading to lower thermoelectric conversion efficiency and higher thermal conductivity [17]. However, doping level should be suppressed in order to achieve higher TE as well as photovoltaic performance. It is also reported that the doping level of GeTe should be in the range of $10^{19}$ cm$^{-3}$ for achieving higher thermoelectric power [18]. The hole concentration of GeTe can be suppressed by doping excess Ge, Bi, In, Zn, Pb, and Sb [17, 19-20]. On the other hand, the carrier type can be alter i.e. n-type conductivity can be achieved by doping GeTe with Co, alloying with AgBiSe$_2$, and Sb doping in GeTe-PbTe alloy [21-23].

However, GeTe semiconductor could be used in TPV cell as it has excellent optical and thermoelectrical properties, for an instance its absorption coefficient is above $2.7\times10^4$ cm$^{-1}$ and resistivity is approximately $2\times10^{-4}$ Ω.cm [24]. GeTe also exhibits a dielectric constant of 40 [6]. In addition, the bulk lifetimes of germanium based material should be between 30 ns to 500 μs [25]. Besides, GeTe is a durable material and due to its durability it can be recovered from any deformation. Moreover, its thermal resistance is high and cost is low [26]. Moreover, its carrier type and carrier concentration can be well controlled by different types of doping in GeTe and its alloys [17, 19-23]. Therefore, these qualities make GeTe a suitable candidate to be used in thermophotovoltaic cells.

In this work, we theoretically calculate the band structure and physical properties of GeTe semiconductor based on density functional theory (DFT) and then explore GeTe as potential candidate in the TPV field utilizing device transport model. These results indicate that GeTe can be an excellent candidate for the application in the thermophotovoltaic cells.

## 2.1 DFT Study on GeTe

### 2.1.1 Computational Methodology

The First-principles calculations on *α*-GeTe were performed in the framework of density functional theory (DFT) utilizing the Cambridge Serial Total Energy Package (CASTEP) programs [27-28]. To investigate the electronic and optical properties of GeTe, the localized density approximation (LDA) of a Perdew Burke Ernzerhof was used as an exchange-correlation function [29]. The Broyden−Fletcher−Goldfarb−Shanno (BFGS) minimization algorithm was applied to optimize the geometry or finding the ground state of GeTe structure. In this geometry optimization, the total energy, maximum force, stress, and displacement were taken into account to $2.0\times10^{-5}$ eV/atom, 0.05 eV/Å, 0.1 GPa, and 0.002 Å, respectively in the lattice as convergence tolerance. The OTFG ultrasoft pseudo-potential framework of Vanderbilt was taken to conceive the interactions between valence electrons and ions. The Koeling-Harmon was applied as relativistic treatment in this study. The plane wave basis set was terminated at 350 eV, and the Brillouin zone was sampled using the 15×15×15 Monkhorst-Pack *k*-point mesh. The high level of 30×30×30 *k*-point mesh was set to calculate the electronic charge density of GeTe. The self-consistent field (SCF) tolerance was fixed to $2.0\times10^{-5}$ eV/atom for this calculation.

## 2.2 Device transport model for TPV cell and calculation

The formula which relates current density (J) with voltage (V) of a pn junction diode under illumination comes out from the principle of classical photovoltaic cell model [30]. It is written in the form of,

$$J = J_{SC} - J_0 \left[\exp\left(\frac{qV}{K_BT}\right) - 1\right] \qquad (1)$$

where, $J_{SC}$, $J_0$, $K_B$, T and q represent the short circuit density, dark current density, Boltzmann constant, cell temperature and charge of electron, respectively.

The short circuit current density, $J_{SC}$ can be calculated using the following formula [31]

$$J_{SC} = \int_0^{\lambda_m} q\varphi(\lambda)IQE(\lambda)d\lambda \qquad (2)$$

In this formula, $\varphi(\lambda)$ is the irradiation spectrum of incident light and $IQE(\lambda)$ denotes the internal quantum efficiency. The product of the electron charge, the photon flux and the internal quantum efficiency is integrated from 0 to the cutting wavelength $\lambda_m$ for reckoning the value of $J_{SC}$.

The blackbody irradiation can be expressed by the following formula,

$$\phi(\lambda) = 2\pi c \lambda^{-4} \left[\exp\left(\frac{hc}{\lambda K_B T_{BB}}\right) - 1\right]^{-1} \tag{3}$$

In the above formula the Plank constant, velocity of light in a free medium and blackbody temperature are presented by the character h, c and $T_{BB}$, respectively.

Total internal quantum efficiency of a simple pn junction is the accumulation of the IQE of the emitter, base and space charge region (scr), respectively. Therefore,

$$IQE = IQE_e + IQE_b + IQE_{scr} \tag{4}$$

The contribution of emitter of the device is expressed by the following equation [32]

$$IQE_e(\alpha) = \frac{\alpha.L_e}{(\alpha.L_e)^2 - 1} \times \left[\frac{S_F + \alpha L_e - (S_F.\cosh\frac{w_e}{L_e} + \sinh\frac{w_e}{L_e})e^{-\alpha.w_e}}{S_F.\sinh\frac{w_e}{L_e} + \cosh\frac{w_e}{L_e}} - \alpha L_e.e^{-\alpha.w_e}\right] \tag{5}$$

The contribution of base can be calculated by the similar way,

$$IQE_b(\alpha) = \frac{\alpha.L_h}{(\alpha.L_h)^2 - 1} \times \left[\alpha L_h - \frac{S_B.\cosh\frac{w_b}{L_h} + \sinh\frac{w_b}{L_h} + (\alpha.L_h - S_B)e^{-\alpha w_b}}{S_B.\sinh\frac{w_b}{L_h} + \cosh\frac{w_b}{L_h}}\right]e^{-\alpha.w} \tag{6}$$

In the equation 5 and 6, the symbol $W_e$ and $W_b$ are the width of emitter and base, $L_e$ and $L_h$ are the minority carrier diffusion length for electrons and holes, $S_F$ and $S_B$ are the front and back surface recombination velocity, respectively.

However, the IQE of space charge region is measured by the product of two probabilities which are a photon approaches the scr and absorb in there.

$$IQE_{scr}(\alpha) = e^{-\alpha.w_e}(1 - e^{-\alpha.w_{scr}}) \tag{7}$$

In the above equation the width of space charge region is denoted by $w_{scr}$.

The space charge region of a simple pn homojunction can be found from the following equation,

$$W_{scr} = \sqrt{\frac{2\varepsilon_o \varepsilon_r V_b (N_a + N_d)}{q(N_a \times N_d)}} \tag{8}$$

where, $N_a$, $N_d$, $V_b$ and $\varepsilon_r$ denote the acceptor concentration, donor concentration, built-in potential and dielectric constant, respectively.

Built-in potential is an essential element for the calculation of the width of the space charge region. This can be calculated from the following formula,

$$V_b = \left(\frac{K_B T}{q}\right) \ln\left(\frac{N_a N_d}{n_i^2}\right) \tag{9}$$

In the above equation, $n_i$ is the intrinsic carrier concentration which can be calculated by the following formula,

$$n_i^2 = \sqrt{N_c N_v} e^{\frac{-E_g}{2K_B T}} \tag{10}$$

In the above equation, $E_g$ is the band gap of the semiconductor, the $N_c$ and $N_v$ are the effective density of states of conduction and valence band, respectively. The $N_c$ and $N_v$ can be calculated by the following formula,

$$N_c = 2\left(\frac{m_e^* K_B T}{2\pi\hbar^2}\right)^{\frac{3}{2}} \tag{11}$$

$$N_v = 2\left(\frac{m_h^* K_B T}{2\pi\hbar^2}\right)^{\frac{3}{2}} \tag{12}$$

where, $m_e^*$ and $m_h^*$ are the effective masses of electrons and holes, respectively.

It should also be noticed that the internal quantum efficiency is strongly depended on the value of absorption coefficient that can be found as follows [31],

$$\alpha(\lambda) = \frac{2\pi\lambda q^2 (2m_r)^{1.5}}{n\varepsilon_0 (cm_0)^2 h^3}\left(h\nu - E_g\right)^{0.5} \langle |P_{cv}|^2 \rangle \tag{13}$$

The conversion mass, $m_r$ is related with the effective masses of electron and hole as,

$$\frac{1}{m_r} = \frac{1}{m_e^*} + \frac{1}{m_h^*} \tag{14}$$

$P_{cv}$ is the optical transition matrix, and we can determine the value of $\langle |P_{cv}|^2 \rangle$ as,

$$\langle |P_{cv}|^2 \rangle = \frac{2m_0^2 E_g (E_g + \Delta)}{3m_e(E_g + 2\frac{\Delta}{3})} \tag{15}$$

where, $\Delta$ is the spin-orbit splitting energy.

The calculation of dark current density can be done by [33],

$$J_0 = qn_i^2 \left( \frac{1}{N_a} \frac{\sqrt{D_e}}{\sqrt{\tau_e}} + \frac{1}{N_d} \frac{\sqrt{D_h}}{\sqrt{\tau_h}} \right) \tag{16}$$

Herein, $D_e$ and $D_h$ denote diffusion coefficient for electron and hole, $\tau_e$ and $\tau_h$ present the electron and hole lifetime, respectively.

The diffusion coefficient for electron and hole can be calculated by the following equations [31],

$$D_e = \frac{K_B \mu_e T}{q} \tag{17}$$

$$D_h = \frac{K_B \mu_h T}{q} \tag{18}$$

where, $\mu_e$ and $\mu_h$ are the mobilities of electron and hole, respectively.

The amount of minority carrier lifetime can be obtained considering the values of Auger recombination coefficient, radiation recombination coefficient and Shockley-Read-Hall (SRH) recombination and can be expressed as [31],

$$1/\tau_e = B_{opt} N_a + 1/\tau_{SRH,e} + AN_a^2 \tag{19}$$

$$1/\tau_h = B_{opt} N_d + 1/\tau_{SRH,h} + AN_d^2 \tag{20}$$

where, $B_{opt}$ and A denote radiative and auger recombination coefficient, respectively. The $\tau_{SRH,e}$ and $\tau_{SRH,h}$ are the SRH recombination lifetime of electron and hole, respectively.

The minority carrier diffusion length for electrons ($L_e$) and holes ($L_h$) can be acquired from the square root of the product of diffusion coefficient and the minority carrier life time [31].

$$L_e = \sqrt{D_e \tau_e} \tag{21}$$

$$L_h = \sqrt{D_h \tau_h} \tag{22}$$

The open circuit voltage, $V_{OC}$ is found from the J-V curve, where the value of current density is zero. It can also be calculated mathematically by the following formula,

$$V_{OC} = \frac{KT}{q}\ln((\frac{J_{sc}}{J_o}) + 1) \tag{23}$$

The ratio of the maximum power point and the product of short circuit current and open circuit voltage is called the fill factor (FF). The identical equation of FF is

$$FF = \frac{P_{max}}{V_{oc}J_{sc}} \tag{24}$$

Finally, the value of efficiency is found from the following equation,

$$\eta = \frac{J_{sc}V_{oc}FF}{\sigma T_{BB}^4} \times 100\% \tag{25}$$

where, σ is the Stefan constant. The value of $\sigma = 5.67 \times 10^{-8}$ Wm$^{-2}$.

GeTe-based thermophotovoltaic device performance was calculated using the above equations at the black body temperature of 1775K and cell temperature of 300K with physical parameters shown in Table 1.

**Table 1:** Physical parameters used for the calculation of PV performance of GeTe TPV cell.

| Parameters | Value | Reference |
|---|---|---|
| Bandgap, $E_g$ | 0.6 eV | [6] |
| Intrinsic carrier concentration, $n_i$ | $2.944 \times 10^{14}$ (cm$^{-3}$) | Calculated |
| Radiation recombination coefficient, $B_{opt}$ | $5.2 \times 10^{-14}$ (cm$^3$.s$^{-1}$) | [34] |
| Auger recombination coefficient, A | $1.1 \times 10^{-31}$ (cm$^6$.s$^{-1}$) | [31] |
| Effective mass of electron, $m_e^*$ | $1.2 m_0$ | [35] |
| Effective mass of holes, $m_h^*$ | $1.15 m_0$ | [36] |
| Spin-orbit splitting energy, $\Delta$ | 0.35 eV | [31] |
| Front surface recombination rate, $S_F$ | $10^5$ cm.s$^{-1}$ | [31] |
| Back surface recombination rate, $S_B$ | $10^5$ cm.s$^{-1}$ | [31] |
| Refractive index, n | 5.5 | [37] |
| Relative dielectric constant, | 40 | [6] |
| Electron mobility, $\mu_e$ | 32.5 (cm$^2$.v$^{-1}$.s$^{-1}$) | [7] |
| Hole mobility $\mu_h$ | 40.7 (cm$^2$.v$^{-1}$.s$^{-1}$) | [7] |

## 3. Results and discussion

### 3.1 First-principles studies on GeTe

#### 3.1.1 Structural properties

The room temperature phase of GeTe is rhombohedral α-phase with a space group of R3m (No. 160) which transform to β-GeTe (space group Fm$\bar{3}$m) at a high temperature of ≈ 650 ± 100K [17, 38]. Therefore, first principle calculation were performed on rhombohedral α-GeTe to visualize its electronic structure and physical properties. Figure 1 depicts the crystal structure of $\alpha$ phase of GeTe. The VESTA software was used to illustrate the GeTe crystal structure. The reported experimental lattice parameters of rhombohedral $\alpha$-GeTe were found to be $a=b=c=$ 4.281 Å, and the angles were also noticed to be $\alpha=\beta=\gamma=$ 58.358° [38]. The atomic positions of Ge (0, 0, and 0) and Te (0.521, 0.521, and 0.521) for this structure were also revealed [38]. The theoretical lattice parameters of $a=b=c=$ 4.22225 Å were achieved in this work with the cell angles of $\alpha=\beta=\gamma=$58.861° following the most effective geometry optimization. Those values are very consistent with the experimental values. The obtained unit cell volume of GeTe was 51.84 Å$^3$ after the compatible geometry optimization. This computation also revealed a 2.83276 Å bond length between Ge and Te.

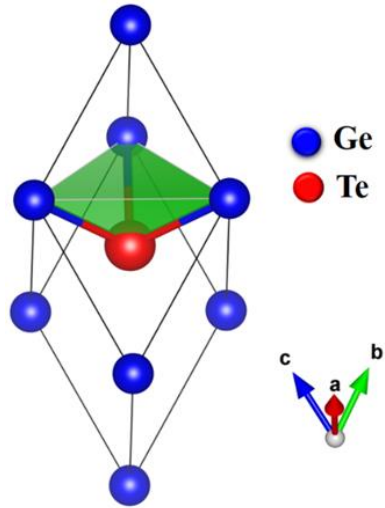

**Figure 1.** Crystal Structure of rhombohedral GeTe semiconductor.

### 3.1.2 Electronic Band Structure and Density of States

The band structure, total density of states (TDOS), partial density of states (PDOS) are involved with the electronic properties of a compound which are also connected to Fermi surface and electronic charge density. The band structure and DOS demonstrate how the eigenvalues of energy fluctuate with the wave vectors in $k$-space providing details on intermolecular and intramolecular bonding [27]. The electronic band structure of $\alpha$-phase of GeTe material is visualized in Figure 2(a). Along the high symmetry of (F-G-Z) directions, the band structure has been examined within the Brillouin zones. The pink color lines indicate the conduction band whereas, violet color lines indicate the valence band. The Fermi energy level, $E_F$ is also indicated by dashed line. A direct bandgap of 0.69 eV at $Z$ point was revealed by this figure which confirms the semiconducting nature of GeTe compound. However, the experimental bandgap for GeTe is 0.60 eV which is very close to our obtained result [6]. This error in bandgap might have occurred because HSE06, a hybrid functional, was not available for use as the exchange correlation function due to the system limitation [27]. Additionally, according to some DFT calculations, GeTe has a direct bandgap of 0.53 eV and an indirect bandgap of 0.38 eV [39]. Therefore, it can be concluded that GeTe with experimental narrow bandgap of 0.6 eV is suitable for the application in TPV field.

Figure 2(b) depicts the total DOS (TDOS) and partial DOS (PDOS) of GeTe semiconducting compound. The TDOS and PDOS are usually studied to investigate the contributions of different orbitals and identify the chemical bonding. The wider curve and sharp peak of DOS profile at Fermi energy level, $E_F$ unveil the structural stability and instability, respectively [21]. Hence, Figure 2(b) clearly indicates the structural stability of GeTe compound. The value of TDOS was seen to ~0.20 states/eV in the figure. This figure also demonstrates that the Ge-4s and Te-5p orbitals are mainly responsible for the TDOS. Then, a strong hybridization between Ge-4s and Te-5p was also noticed at $E_F$. However, Ge-4p and Te-4s have very little contributions in the TDOS of GeTe structure.

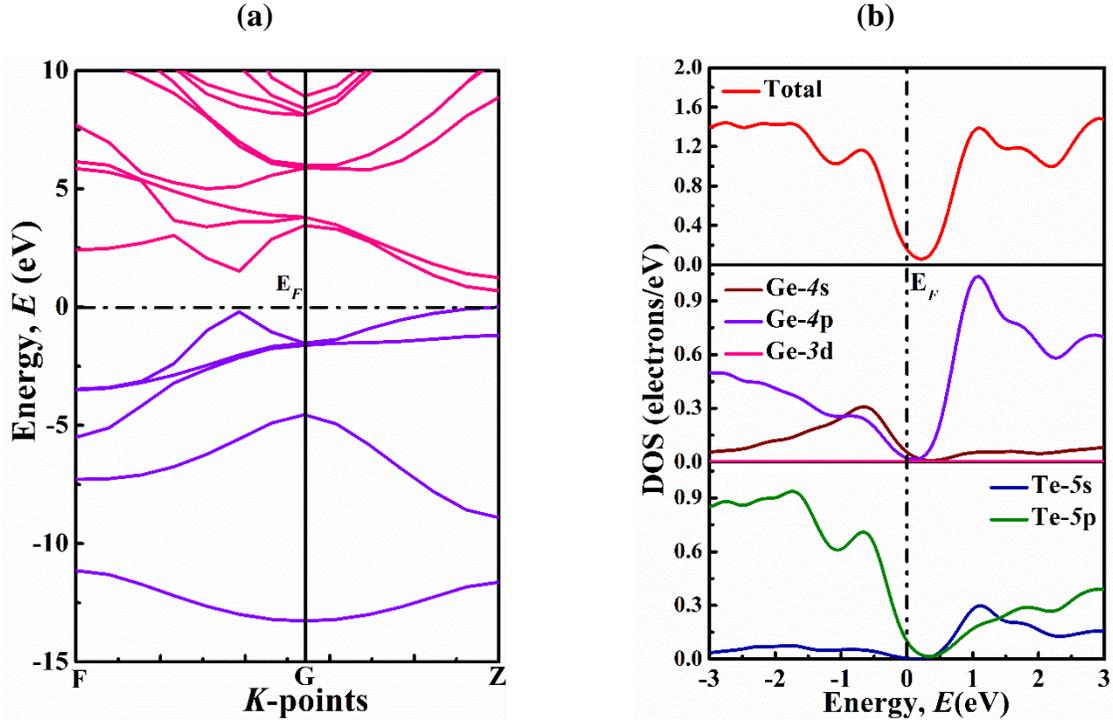

**Figure 2.** The (a) Electronic band structure and (b) DOS of GeTe semiconductor.

### 3.1.3 Electronic Charge Density

Figure 3 represents the charge density mapping of rhombohedral GeTe. The valence electronic charge density contour map indicates the total electronic charge distribution (in the units of e/Å$^3$) of materials [40]. The intensity of the electron concentration is depicted on the right-side scale. Blue color displays a low electron density, while red color displays a high electron density in the scale. The figure shows that the charge distribution is not spherical around all the atoms which indicates covalent nature between them [27]. This demonstrates that the GeTe compound is non-metallic which is also guaranteed by the electronic band structure (Figure 2).

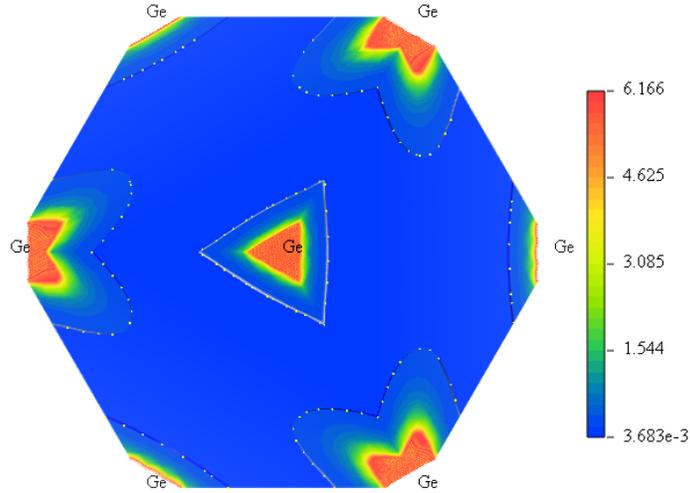

**Figure 3.** Electronic charge density of GeTe semiconductor.

### 3.1.4 Optical Properties

The energy dependent optical properties of GeTe semiconductor has been studied broadly and shown in Figure 4. A number of factors are required to investigate a material's optical properties including the absorption coefficient, dielectric function, refractive index, conductivity, reflectivity, and loss function that depend on the energy of the incident electromagnetic radiation.

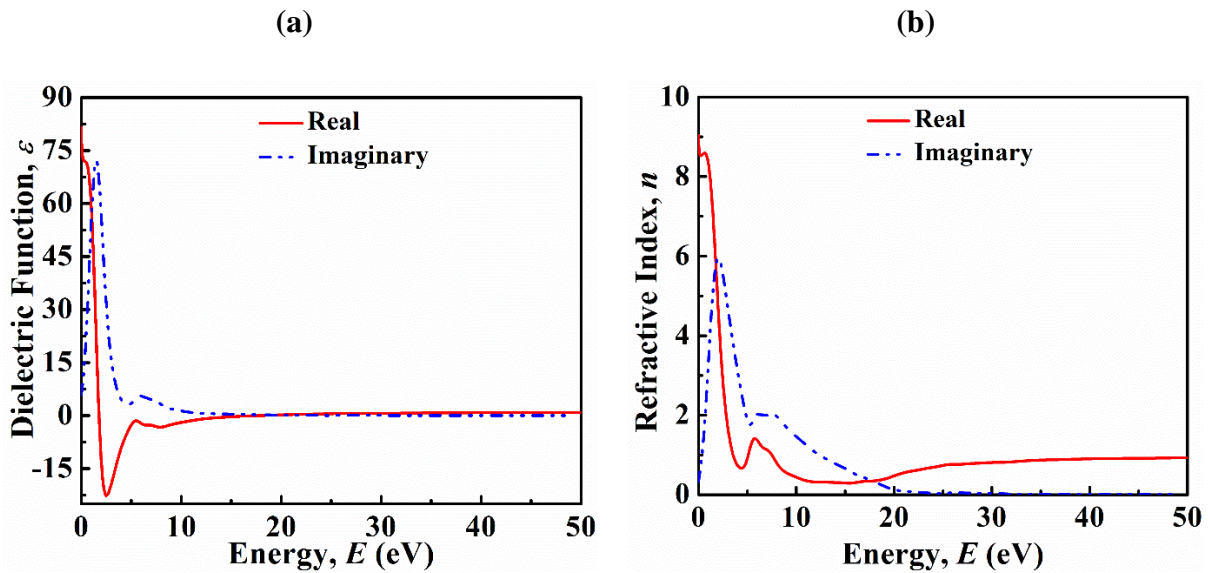

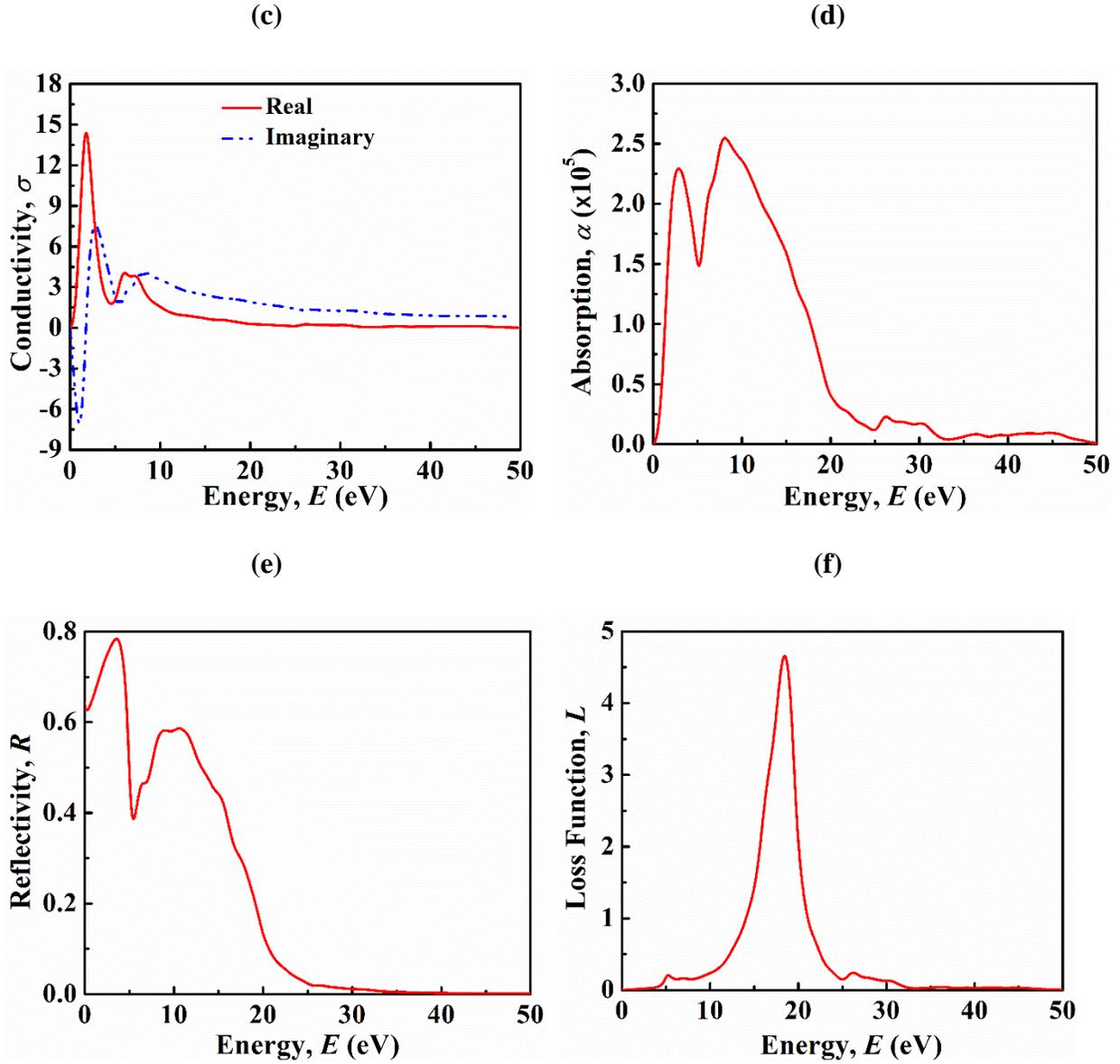

**Figure 4.** Frequency dependent optical properties: (a) Dielectric Function, (b) Refractive index, (c) Conductivity, (d) Absorption, (e) Reflectivity, and (f) Loss function of GeTe semiconductor.

The real ($\varepsilon_1$) and imaginary ($\varepsilon_2$) part of dielectric function of GeTe are presented in Figure 4(a). The dielectric function gives an explanation of a material's response to incident photons as a function of energy. The static dielectric constant is the real part of dielectric constant that exists at absolute zero frequency. The value of static dielectric constant for GeTe is found to be 81.58. The high peak of real portion might be originated due to the intraband transitions of conduction electrons [40]. The real dielectric function becomes zero after 2.0 eV energy and it remains almost zero in the entire energy range. The imaginary dielectric constant indicates the maximum value of 72 at 1.5 eV and also shows the same behavior as real portion after 10 eV energy.

The refractive index is a crucial characteristics of a material which quantifies how much light is refracted. The calculated refractive index spectra for GeTe is illustrated in Figure 4(b). The static refractive index found in the figure is 9.09. The refractive index changes depending on the energy, demonstrating the photorefractive property of the GeTe [27].

The optical conductivity ($\sigma$) spectra of GeTe is visualized in the Figure 4(c). The true portion of photoconductivity does not begin with the zero photon energy, which denotes the GeTe material having bandgap energy between the conduction and valence band [40]. The electronic band structure of GeTe in the Figure 2 also ensures that. The maximum photoconductivity of 14.40 for GeTe structure was obtained at 1.9 eV.

Figure 4(d) shows the absorption coefficient ($\alpha$) for the GeTe compound. It is seen in the figure that due to semiconducting nature and the calculated band gap of 0.69 eV, which is familiar with absorption edges as well, absorption for GeTe starts ascending at about 0.70 eV. Moreover, the low-energy infrared region of the spectra results from the intraband transition and, the peaks in the high-energy region of absorption spectra could possibly be the result of the interband transition [40]. However, this high absorption spectra of GeTe material indicates that it can be used as absorber for solar energy conversion.

The reflectivity spectra of GeTe is demonstrated in the Figure 4(e). This figure indicates the high conductance characteristics of GeTe due to the high reflectance value in low energy region [27]. At low energies, the GeTe has a maximum reflectance of 78% of total radiation.

The loss function spectra of GeTe is also displayed in Figure 4(f). The energy loss function, a crucial index, can be used to indicate how much energy is lost as an electron moves swiftly through a material [41]. The plasma resonance is shown by the largest peak in the figure, and the frequency corresponding with it is referred to as the plasma frequency ($\omega_p$) [27]. The highest value of loss function for the GeTe semiconductor is observed at 18.48 eV. When incident photons surpasses plasma frequency, the GeTe semiconductor becomes transparent.

## 3.2 TPV performance of GeTe

In this section, we discuss about the performance of GeTe thermophotovoltaic cell. Figure 5(a) illustrates the schematic diagram of the proposed n-GeTe/p-GeTe homojunction TPV solar cell that can be grown on Si substrate [15,16]. Herein, n-type GeTe functions as the emitter layer and p-type GeTe works as the absorber layer. GeTe semiconductor possesses an electron affinity of 4.8 eV [24] together with the ionization energy of 5.4 eV. The energy band diagram of the TPV cell with np structure has been depicted by the Figure 5(b). The energy levels of conduction and valence band are denoted by $E_c$ and $E_v$, respectively. Furthermore, the path of electrons and holes have also be presented by the blue arrows. The flat-band energies for anode and cathode were calculated to be 5.5 and 4.6 eV using a one-dimensional solar cell capacitance (SCAPS-1D) software developed by M. Burgelman et al. at the University of Gent, Belgium [42]. Therefore, metal with work function≥5.5 eV e.g. Pd, Pt etc. and metal with work function ≤4.6 eV e.g. Al, Ag, Ti etc. can used as suitable anode and cathode, respectively.

The photovoltaic performances of np GeTe TPV cell was calculated solving the equations of device transport model considering the width of the emitter and base of 0.1 and 1µm, respectively. The optimized value of both the donor and acceptor density was set to $10^{19}$ cm$^{-3}$.

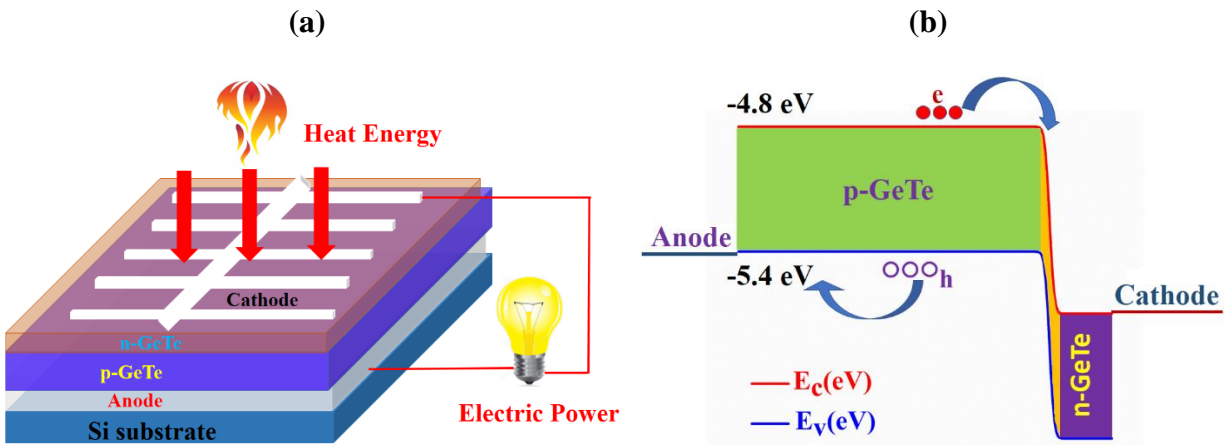

**Figure 5:** The (a) schematic block and (b) energy band diagram (drawn not to scale) of GeTe pn junction TPV cell.

### 3.2.1 Impact of base on GeTe TPV performance

The Figure 6(a) exhibits the change of photovoltaic parameters with the variation of thickness of base of the GeTe TPV cell. There is a rapid change of all parameters with the width of the base absorber layer up to 2.5 µm and beyond this value all of them have labelled off. The short circuit current, $J_{SC}$ has incremented from 12.6 to 21 A/cm² with the thickness of the base because of the production of more charge carriers in the device [42]. And after attaining this value the current was almost saturated. The open circuit voltage, $V_{OC}$ of the device improves from 360 to 370 mV with thickness of the base of the TPV device. However, only a tiny advancement of 0.5% has been noticed for the fill factor (FF) that was increased from 74.5 to 75%. The efficiency, η of the TPV device significantly increased from 7 to 11% due to the increase in $J_{SC}$ and $V_{OC}$ of the GeTe TPV cell. However, thickness of the base was considered to be 1 μm since the cost of the device would be higher for thicker base of the device.

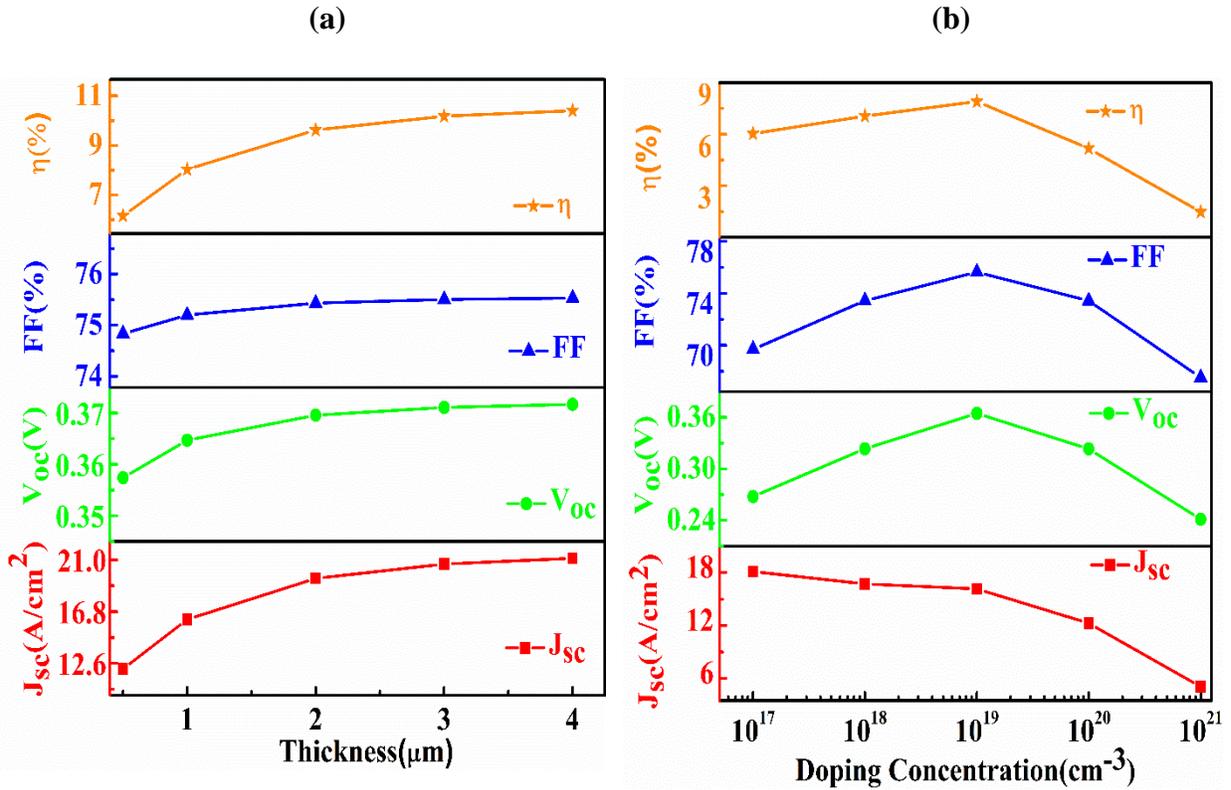

**Figure 6.** Variation of output parameters as a function (a) thickness, (b) doping concentration of base of the GeTe TPV cell.

The variations of PV parameters with respect to doping concentration of the base of the GeTe TPV cell has been illustrated in Figure 6(b). All the performance parameters have deteriorated with the higher doping concentration as because of the radiaitve recombination loss in the device [42]. The value of $J_{SC}$ has plummeted from 18 to almost zero in the observed range of doping concentration. Both the $V_{OC}$ and the FF was depictured a moderate positive change up to $10^{19}$ cm$^{-3}$, which was 270 to 360 mV and 70 to 76%, respectively. The increase in $V_{OC}$ and hence FF was resulted due to the increase in built-in potential with doping concentration [43-44]. In the doping range of $10^{17}$ to $10^{19}$ cm$^{-3}$, there was a 2% advancement of efficiency and it was increased from 6 to 8%. However, above this doping concentration, all the output parameters were likely to decrease for the sake of higher chance of recombination at a great value of carrier concentration. At higher carrier, the impurity scattering and carrier recombination rate is enhanced which deteriorates the device performance [42, 45].

The device efficiency was found to be 7.9% with $J_{SC}$=16.16 A/cm$^2$, $V_{OC}$=0.360 V and FF=75.51%, respectively for the base thickness and doping of 1 μm and $10^{19}$ cm$^{-3}$, respectively.

### 3.2.2 Effects of emitter on GeTe TPV performance

Figure 7(a) illustrates the influence of the thickness of emitter of GeTe TPV cell on four parameters called short circuit current density, open circuit voltage, fill factor and efficiency, respectively. From the figure, it is lucidly noticed that there is a negative impact of thickness of emitter on $J_{SC}$. The grade of $J_{SC}$ has decreased from approximately 16.8 to 14.4 A/cm$^2$. The reason behind this reduction is, at short wavelength, the internal quantum efficiency is highly affected by the thickness of emitter. High depth of emitter is responsible for low production of IQE which is related to the built-in potential that can affect the minority carriers' movement [33]. An avoidable change has been noticed for both the $V_{OC}$ and the fill factor of the device. Due to the reduction of $J_{SC}$ the efficiency has showed a meager negative change from 8.5 to 7% when thickness of the emitter increased from 0.05 to 0.3 µm. However, thickness of the emitter was considered as 0.1 µm as the film quality degrades when the thickness becomes very thin.

The domination of doping concentration of emitter on the four output parameters has been presented in Figure 7(b). As the minority carrier lifetime reduces with a higher doping concentration, the value of $J_{SC}$ has depicted a wane from 18 to 15 A/cm$^2$ [46]. However, the other three parameters have showed an advancement with the improvement of carrier density up to $10^{19}$ cm$^{-3}$. After reaching the peak value, all the three parameters $V_{OC}$, FF and efficiency illustrated a moderate decrement because of the increment of recombination at a higher doping concentration. For doping concentration up to $10^{19}$ cm$^{-3}$, the $V_{OC}$, FF and efficiency have improved from 260 to 350 mV, 69 to 76 % and 6 to 8.5 %, respectively and above this value all of them have showed moderate downward change from their apex value to their initial value.

The GeTe TPV cell showed an efficiency of 7.9% for the emitter thickness and doping of 0.1 µm and $10^{19}$ cm$^{-3}$, respectively.

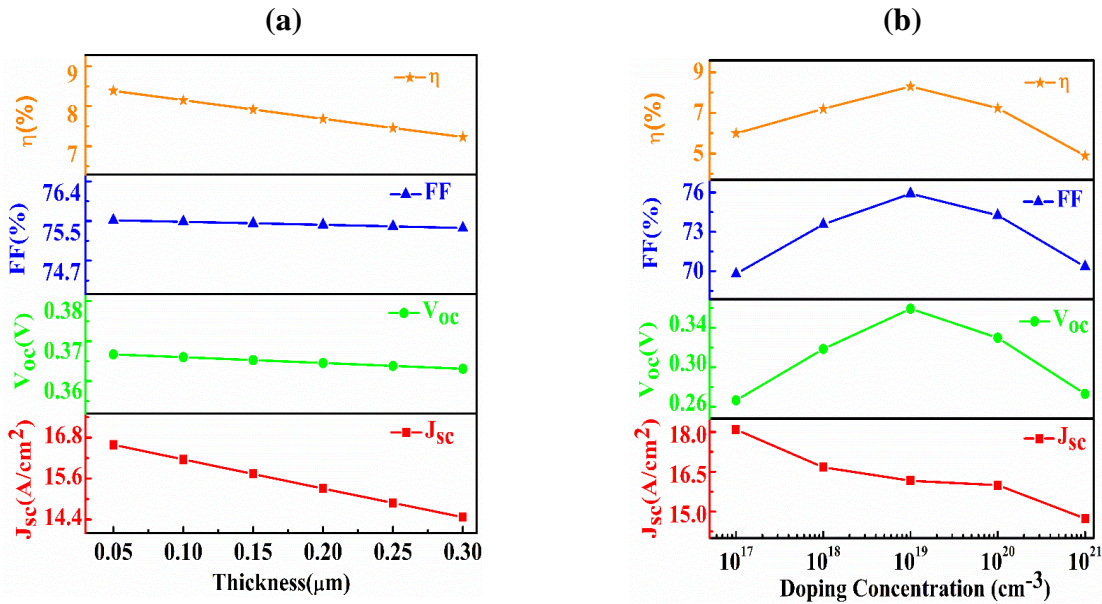

**Figure 7.** Variation of output parameters as a function of (a) thickness, and (b) doping concentration of emitter of the GeTe TPV cell.

### 3.2.3 Impact of black body temperature on GeTe TPV performance

Figure 8(a) delineates the variation of photon flux with respect to wavelength at different black body ($T_{BB}$) temperature. A noticeable advancement of photon flux and the left shift of radiation peak have been observed with the increasing $T_{BB}$. At the black body temperature of 1775 K, GeTe TPV cell with cutoff wavelength of 2070 nm can capture maximum thermal radiation of photons, which has proved GeTe material as a potential candidate for a TPV cell.

The variation of absorption coefficient and IQE with wavelength have been showed by the line plot in Figure 8(b). It is noticed that the value of absorption coefficient first stays almost constant with wavelength up to 2000 nm then it plunged and at the cut off wavelength it falls to approximately zero value. The order of the absorption coefficient is of the order of $10^4$ cm$^{-3}$ which is consistent with Figure 4(d) calculated from first-principles study. The behavior of absorption coefficient with the variation of wavelength can be depicted by the equation $\alpha h\nu = A(h\nu - E_g)^p$, where A is a constant, p is an exponent [47]. Nonetheless, IQE has maintained almost constant value of just over 60% up to 2000 nm of wavelength, before falling to zero value at the cut off wavelength of 2070 nm.

Figure 8(c) represents the strong dependence of the output parameters of GeTe thermophotovoltaic cell on the blackbody temperature. All of the parameters have depictured a great development with the increase of the blackbody temperature. At a black body temperature of 1000K, $J_{SC}$ is almost zero, but at 2500K $J_{SC}$ stands on 87 A/cm$^2$. This is resulted because the photon flux increases with the enlargement of $T_{BB}$ which significantly increases the short circuit current in the device. A rapid change of $V_{OC}$ from 280 to 420 mV has been noticed in the observed range of temperature. As the $J_{SC}$ significantly increases the $V_{OC}$ of the GeTe TPV device also increases. The number of carrier collection is increased with $T_{BB}$ which leads to the band filling effect. Due to the improvement of maximum power, FF depicts a prosperity from 69 to 77%. The upliftment of these three parameters forces the efficiency of the TPV cell to reach from 0 to about 15%.

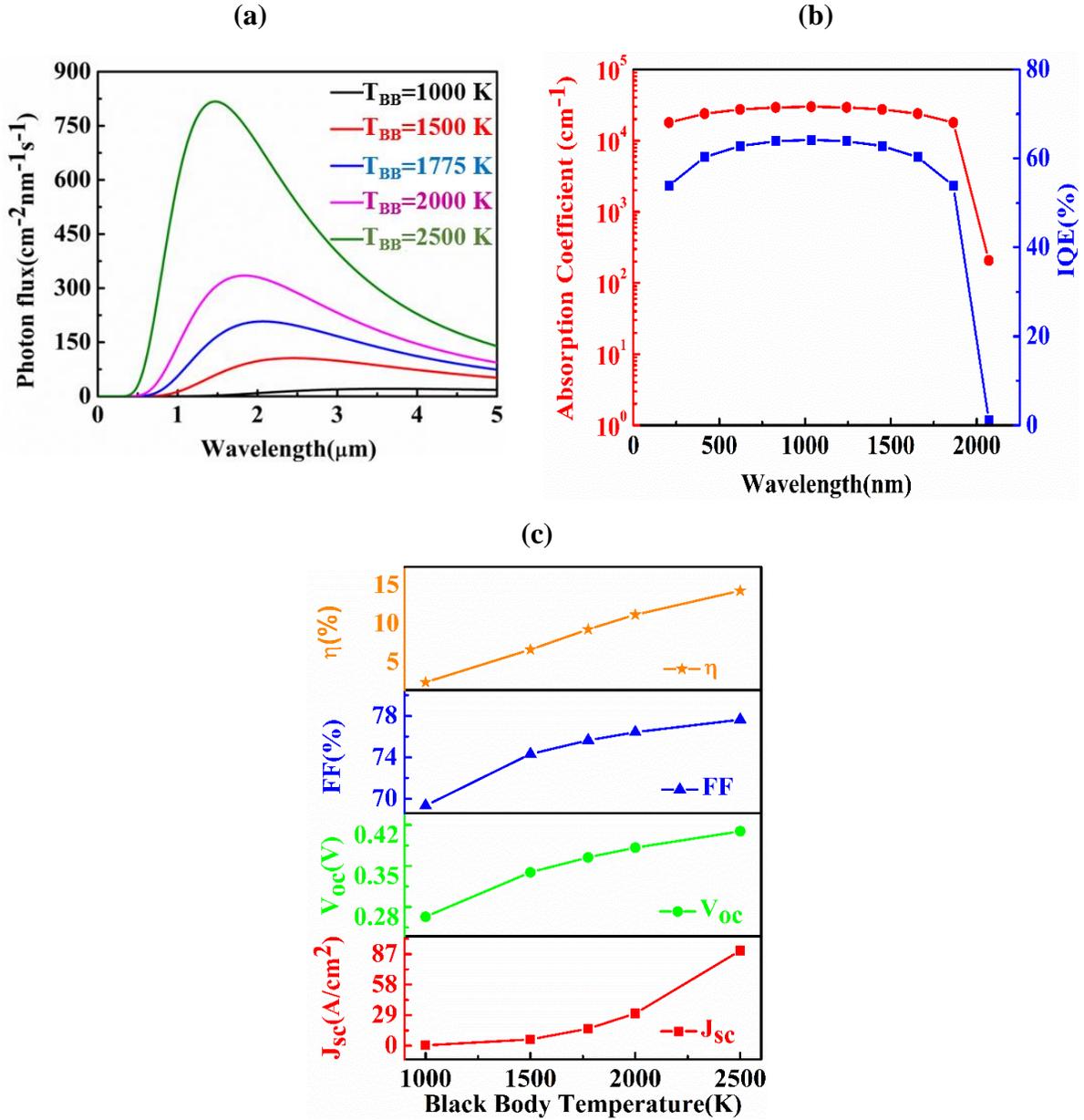

**Figure 8.** Variation of (a) photon flux as a function of wavelength at different temperature, (b) absorption coefficient and IQE as a function of wavelength, and (c) output PV parameters as a function of black body temperature.

### 3.5.6. Impact of carrier Lifetime on GeTe performance

Figure 9(a) presents the fluctuation of output parameters of GeTe TPV cell as a function of the life time of holes. From the figure, it is cleared that the PV parameters are very sensitive with hole life time. The life time was varied from 1 to 10000 ns and over this range $J_{SC}$ plummeted from 12.6 to 16.8 A/cm$^2$. The reason of this improvement is the enhancement of mean free path with the life time, which is directly related to the internal quantum efficiency of the device. There is a moderate amelioration of open circuit voltage from 330 to 390 mV due to the reduction of dark current. Both the FF and efficiency increases with hole lifetime. The FF displays an elevation from 73 to 76.5% whereas the efficiency increases from 5 to 8.5% in turn. The enrichment of $J_{SC}$ and $V_{OC}$ have pulled them on a better position.

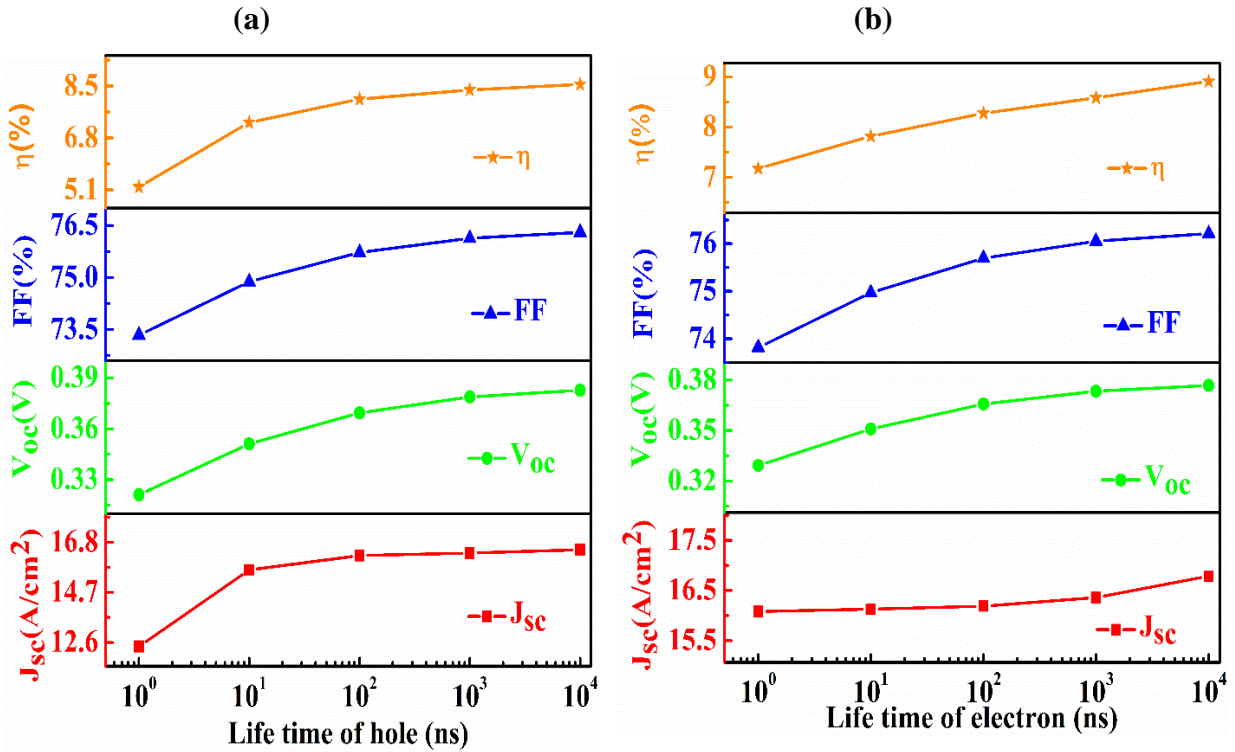

**Figure 9.** Variation of output parameters as a function of (a) holes lifetime, and (b) electrons lifetime of GeTe TPV ccell.

Domination of electron lifetime over the PV parameters are shown in Figure 9(b). The lifetime was varied from 1 to 10000 ns. Over this huge range, $J_{SC}$ exhibited only a minute change from 16 to 16.5 A/cm$^2$. Nevertheless, $V_{OC}$ delineates a moderate change of almost 60 mV (from 320 to 380 mV) for the sake of the mitigation of dark current. The fill factor of the device increases by approximately 2% which increases from 74% to 76% in the observed empirical range. At the same time, the value of efficiency increases from 7 to 9 % as the other parameters have been ameliorated with the electron lifetime.

## 4. Conclusion

First-principles study was carried out to calculate the band structure and physical properties of GeTe semiconductor to visualize its potential in thermophotovoltaic application. The direct bandgap of the semiconductor was calculated to be 0.69 eV which is close to the experimental value. The DOS and absorption coefficient were also calculated. The GeTe TPV cell was modeled and PV performance of the device was calculated using transport model with simple np homojunction structure. The optimized emitter and base thicknesses were 0.1 and 1 μm, respectively. Doping order for both the emitter and base was $10^{19}$ cm$^{-3}$. The optimized device shows an efficiency of 7.9%. These results indicate that GeTe could be a promising semiconductor for the application in TPV cell in future.

**Corresponding author:**

*E-mail: jak_apee@ru.ac.bd (Jaker Hossain).

**NOTES:** The authors declare no competing financial interest.